\documentclass[12pt]{article}

\begin{document}

\Large
\begin{center}

   \vspace*{3ex}
     {\bf Q-deformation, discrete time and quantum information as fiber space  } \\
             \vspace*{2ex}
              {\bf Jaroslav Hruby}

   {\bf Institute of Physics AV CR,Czech Republic }
  {\normalsize \it e-mail: hruby.jar@centrum.cz  }


\vspace*{2ex}
\begin{abstract}

     In this paper we show the connection between the q-deformation and discrete
     time, starting from the q-deformed Heisenberg uncertainty relation and q-deformation calculus.
     We show that time has discrete nature and for this case we construct the connection between
     quantum information and spacetime via fiber space structure.

\end{abstract}

\end{center}

\vspace*{3ex}

\normalsize

\section{Heisenberg uncertainty relation, q-deformation and discretization of spacetime}

\setcounter{equation}{0}
   Following the papers \cite{1} let observables $Q$ and $P$ fulfill:
\begin{equation}
   {\bf a} = 1/\sqrt{2}({\bf Q} + i\,{\bf P}) \ , \qquad
   {\bf a}^{+} = 1/\sqrt{2}({\bf Q} - i\,{\bf P})            \label{1.1}
\end{equation}

where ${\bf a}$ and ${\bf a}^{+}$ are creation and anihilation
operators as usual and \([{\bf Q},{\bf P}]=i\).

   We shall define $\bigtriangleup P \equiv P - \langle P\rangle$
and $\bigtriangleup Q \equiv Q  -  \langle Q\rangle$  . Then the
famous Heisenberg uncertainty relation for observables $Q$ and $P$
:
\begin{eqnarray}
 \frac{1}{2} |\langle[{\bf Q},{\bf P}]\rangle|
   = \frac{1}{2}
     |\langle[\bigtriangleup {\bf Q},\bigtriangleup {\bf P}]\rangle|
                                                   \nonumber\\
   \leq \langle\bigtriangleup {\bf Q}^{2}\rangle^{\frac{1}{2}}
      \langle\bigtriangleup {\bf P}^{2}\rangle^{\frac{1}{2}}  \ .
           \label{1.2}
\end{eqnarray}

The  uncertainty can be understood as the product $\Delta Q \Delta
P$ of standard deviations of two observables for a system in a
given state and it is often used to identify it with
noncommutativity of quantum observables under consideration.

If we shall assume the q-deformation of the commutator between
creation and anihilation operator we can write :
\begin{equation}
  {\bf a}\,{\bf a}^{+} - q{\bf a}^{+}{\bf a} = I \ ,     \label{1.3}
\end{equation}

where $q$ is the deformation parameter $0<q\leq 1$ and ${\bf I}$
is identity operator.

   Let {\bf P} and {\bf Q} are Hermitean operators, which via
 {\bf a} and {\bf a}$^{+}$ have form:
\begin{equation}
  {\bf P} = \alpha {\bf a}+\alpha^{\ast}{\bf a}^{+} \ , \qquad
  {\bf Q} = \beta {\bf a}+\beta^{\ast}{\bf a}^{+} \ ,       \label{1.4}
\end{equation}

where $\alpha,\ \beta$ are complex parameters.

   Then from q-commutation relation (\ref{1.3}) follows:
\begin{equation}
   [{\bf P},{\bf Q}] = (\alpha\beta^{\ast}-\alpha^{\ast}\beta)
     [{\bf I}+(q-1){\bf a}^{+}{\bf a}] = {\bf R}  \ .     \label{1.5}
\end{equation}

   We can see for \(q=1\) and
$\alpha(\beta)^{\ast}-(\alpha)^{\ast}\beta= - i$ that (\ref{1.5})
are ordinary uncertainty relation.

   Uncertainty relation follows directly from (\ref{1.5})
\begin{equation}
   \frac{1}{4}\,|\langle {\bf R}\rangle|^{2} \leq
     \langle\bigtriangleup {\bf Q}^{2}\rangle
       \langle\bigtriangleup {\bf P}^{2}\rangle \ ,          \label{1.6}
\end{equation}

 and it is the known form for operators fulfilling
 $[{\bf Q},{\bf P}]=i{\bf R}$.

    For operators Q,P,R  the q-deformed uncertainty relation are:
\begin{eqnarray}
   \langle\bigtriangleup {\bf Q}^{2}\rangle =
       |\beta|^{2}[1+(q-1)|\langle {\bf a}\rangle|^{2}] \ ,     \label{1.7}   \\
   \langle\bigtriangleup {\bf P}^{2}\rangle =
       |\alpha|^{2}[1+(q-1)|\langle {\bf a}\rangle|^{2}] \ ,     \label{1.8}   \\
    \langle {\bf PQ}\rangle-\langle {\bf QP}\rangle =
       (\alpha\beta^{\ast}-\beta^{\ast}\alpha)
          [1+(q-1)|\langle {\bf a}\rangle|^{2}]  \ .           \label{1.9}
\end{eqnarray}

     It is important that (\ref{1.7})--(\ref{1.9}) are valid' for
     arbitrary operators which are fulfilling(\ref{1.6}) and representing conjugate physical observables.

It was demonstrated that the more basic notions of expected value,
variance and uncertainty relation also have a clear geometric
interpretation.This interpretation is based directly on the
association of observables with vector fields on the sphere of
states and does not employ the Hamiltonian formalism on the phase
space. This makes the interpretation particularly transparent and
naturally leads one to a geometric uncertainty identity. Here one
is faced then  a new point of view on quantum mechanics (QM) that
makes that theory quite similar to Einstein's general relativity.

It is also well known that uncertainty relation in q-QM for
coordinate $x$ and $p$ can be obtain via the  way :

     let Bargmann-Fock`s operators have the form
\begin{eqnarray}
 a & = & \frac{1}{2L}\,x-\frac{i}{2K}\,p \ , \label{1.10} \\
 a^{+} & = & \frac{1}{2L}\,x+\frac{i}{2K}\,p=\partial_{a} \ ,
                              \label{1.11}
\end{eqnarray}

and $q$-commutator
\begin{equation}
 a\,a^{+}- {q_{0}}^2\,a\,a^{+}=[a,\partial_{a}]_{q_{0}}=1        \label{1.12}
\end{equation}
wher $q_{0}$ is real deformation parameter connected with the
constants $K$ and $L$ via \(K L =\frac{\hbar}{4}\,(q_{0}^{2}+1)\)
\@.
    Here the constants $L$ and $K$ have the dimension of length and impuls.

    The commutation relation between $x$ a $p$ is

\begin{equation}
 [x,p]=i\,\hbar\,(1+f(q_{0},x,p)) \ ,                   \label{1.13}
\end{equation}
where
\begin{equation}
 f(q_{0},x,p)=\frac{{q_{0}}^2-1}{4}\left(\frac{x^{2}}{L^{2}}+
     \frac{p^{2}}{K^{2}}\right) \ .                      \label{1.14}
\end{equation}

   Then the q-deformed Heisenberg uncertainty relation follows
\begin{equation}
 \Delta x\,\Delta p \geq \frac{\hbar}{2}\,\Bigl[ 1+f(q_{0},
  (\Delta x)^{2}+\langle x\rangle^{2},
  (\Delta p)^{2}+\langle p\rangle^{2})\Bigr] \ .       \label{1.15}
\end{equation}

In every case of q-deformed QM we have minimal uncertainty in $x$
and also in $p$, which are for $q_{0}>1$ :
\begin{equation}
 \Delta x_{0}=L\sqrt{1-q_{0}^{-2}} \ , \qquad
 \Delta p_{0}=K\sqrt{1-q_{0}^{-2}} \ .       \label{1.16}
\end{equation}

   It gives the way to the discretization of the spacetime in
   q-deformed world.

\section{Q-deformation, discrete time and quantum information}
   Let us suppose that $q_{E}$ is the parameter of the
discretization of spacetime.

   Let us consider the discretization of standard differential calculus in one
space dimension
\begin{equation}
   [x,dx] = dxq_{E}  ,
      \label{2.1}
\end{equation}
and the action of the discrete translation group
\begin{equation}
   x^ndx = dx(x+q_{E})^n  ,
      \label{2.2}
\end{equation}
\begin{equation}
  \psi(x)dx = dx\psi(x+q_{E}) ,
      \label{2.3}
\end{equation}
for any wave function $\psi$ of the Hilbert space of
QM with the discrete space variable.   \\

   The discrete space variable is defined as $x=nq_{E}$, where n is an integer and
and violation parameter $q_{E}$ is the interval between two
discrete space points in this space variable.

   If we define the derivatives by
\begin{equation}
  d\psi(x) = dx(\partial_{x}\psi)(x) =
  (\stackrel{\leftarrow}{\partial}\psi)(x)dx,
      \label{2.4}
\end{equation}
\begin{equation}
  (\partial_{x}\psi)(x) = \frac{1}{q_{E}}[\psi(x+q_{E})-\psi(x)],
      \label{2.5}
\end{equation}
\begin{equation}
  (\stackrel{\leftarrow}{\partial_{x}}\psi)(x) = \frac{1}{q_{E}}[\psi(x)- \psi(x-q_{E})],
      \label{2.6}
\end{equation}
\begin{equation}
(\stackrel{\leftarrow}{\partial_{x}}\psi)(x) =
(\partial_{x}\psi)(x-q_{E})  ,
      \label{2.7}
\end{equation}
then the ordinary one-dimensional  Schr$\ddot{o}$dinger equation
will be
\begin{equation}
 \frac{1}{2}\frac{d^2\psi(x)}{dx^2} + [E - U(x)]\psi(x)  = 0,
      \label{2.8}
\end{equation}
with the potential $U(x)$ and wavefunction
$\psi(x)\equiv\psi(E,x)$, corresponding to energy value E, has on
the discrete space the form
\begin{equation}
 \frac{1}{2l^2}[\psi((n+1)q_{E})-2\psi(nq_{E})+ \psi((n-1)q_{E})]+ [E - U(nq_{E})]\psi(nq_{E})  = 0.
      \label{2.9}
\end{equation}

We now show the coincidence between such discretization model,
noncommutative differential calculus and q-deformed QM, assuming
$q^2\approx 1$.

Let us suppose that ordinary continuum space variable y in QM has
the form:
\begin{equation}
y = \lim_{q_{E}\rightarrow0}(1+q_{E})^{\frac{x}{q_{E}}}=
  e^{x}.
\label{2.10}
\end{equation}

   Using Eqs.(2.4-2.7) and (2.10) we get:
\begin{equation}
\partial_{y} = y^{-1}\partial_{x} = (q_{E} + 1)^\frac{-1}{q_{E}}\partial_{x}
\label{2.11}
\end{equation}
   Thus, using $q_{E}\equiv q^2-1$, we have
\begin{equation}
  (\partial_{y}\psi)(y) = \frac{\psi((q_{E}+1)y)-\psi(y)}{q_{E}y}=
         \frac{\psi(q^2y)-\psi(y)}{(q^2-1)y}
      \label{2.12}
\end{equation}
\begin{equation}
(\stackrel{\leftarrow}{\partial_{y}}\psi)(y)   = (q_{E}+1)
\frac{\psi(y)-\psi((q_{E}+1)y))}{q_{E}y} =
\frac{\psi(y)-\psi(q^2y)}{(1-q^{-2})y} \\      \label{6.13}
\end{equation}
what represents derivatives in the differential on the quantum
hyperplane.

   We can see that for $q_{E}=0$ or $q^2=1$ we have the ordinary QM and continuous
space-time.

\section{ Q-deformation calculus, non-commutativity  and differential geometry}
Aspects of gauge theory, Hamiltonian mechanics and QM arise
naturally in the mathematics of a non-commutative framework for
calculus and differential geometry.

Following summary paper \cite{2} we can see that the q-deformation
calculus has the deep connection with differential geometry and
gauge fields.

There is shown the constructions of the non-commutativity are
performed in a Lie algebra $\cal A.$ One may take $\cal A$ to be a
specific matrix Lie algebra, or abstract Lie algebra. If $\cal A$
is taken to be an abstract Lie algebra, then it is convenient to
use the universal enveloping algebra so that the Lie product can
be expressed as a commutator. In making general constructions of
operators satisfying certain relations, it is understood that one
can always begin with a free algebra and make a quotient algebra
where the relations are satisfied. \bigbreak

On $\cal A,$ a variant of calculus  is built by defining
derivations as commutators (or more generally as Lie products).
For a fixed $N$ in $\cal A$ one defines
$$\nabla_N : \cal A \longrightarrow \cal A$$ by the formula
$$\nabla_{N} F = [F, N] = FN - NF.$$
$\nabla_N$ is a derivation satisfying the Leibniz rule.
$$\nabla_{N}(FG) = \nabla_{N}(F)G + F\nabla_{N}(G).$$
\bigbreak

There are many motivations for replacing derivatives by
commutators. If $f(x)$ denotes (say) a function of a real variable
$x,$ and $\tilde{f}(x) = f(x+h)$ for a fixed increment $h,$ define
the {\em discrete derivative} $Df$ by the formula $Df = (\tilde{f}
- f)/h,$ and find that the Leibniz rule is not satisfied. One has
the basic formula for the discrete derivative of a product:
$$D(fg) = D(f)g + \tilde{f}D(g).$$ Correct this deviation from the
Leibniz rule by introducing a new non-commutative operator $J$
with the property that
$$fJ = J\tilde{f}.$$ Define a new discrete derivative in an extended non-commutative algebra by the formula
$$\nabla(f) = JD(f).$$ It follows at once that
$$\nabla(fg) = JD(f)g + J\tilde{f}D(g) = JD(f)g + fJD(g) = \nabla(f)g + f\nabla(g).$$
Note that $$\nabla(f) = (J\tilde{f} - Jf)/h = (fJ-Jf)/h = [f,
J/h].$$ In the extended algebra, discrete derivatives are
represented by commutators, and satisfy the Leibniz rule. One can
regard discrete calculus as a subset of non-commutative calculus
based on commutators. \bigbreak

In $\cal A$ there are as many derivations as there are elements of
the algebra, and these derivations behave quite wildly with
respect to one another. If one takes the concept of {\em
curvature} as the non-commutation of derivations, then $\cal A$ is
a highly curved world indeed. Within $\cal A$ one can build a tame
world of derivations that mimics the behaviour of flat coordinates
in Euclidean space. The description of the structure of $\cal A$
with respect to these flat coordinates contains many of the
equations and patterns of mathematical physics. \bigbreak

\noindent The flat coordinates $X_i$ satisfy the equations below
with the $P_j$ chosen to represent differentiation with respect to
$X_j.$:

$$[X_{i}, X_{j}] = 0$$
$$[P_{i},P_{j}]=0$$
$$[X_{i},P_{j}] = \delta_{ij}.$$
Derivatives are represented by commutators.
$$\partial_{i}F = \partial F/\partial X_{i} = [F, P_{i}],$$
$$\hat{\partial_{i}}F = \partial F/\partial P_{i} = [X_{i},F].$$
Temporal derivative is represented by commutation with a special
(Hamiltonian) element $H$ of the algebra:
$$dF/dt = [F, H].$$
(For quantum mechanics, take $i\hbar dA/dt = [A, H].$) These
non-commutative coordinates are the simplest flat set of
coordinates for description of temporal phenomena in a
non-commutative world. Note:

\noindent {\bf Hamilton's Equations.} $$dP_{i}/dt = [P_{i}, H] =
-[H, P_{i}] = -\partial H/\partial X_{i}$$
$$dX_{i}/dt = [X_{i}, H] = \partial H/\partial P_{i}.$$
These are exactly Hamilton's equations of motion. The pattern of
Hamilton's equations is built into the system. \bigbreak

\noindent {\bf Discrete Measurement.} Consider a time series $\{X,
X', X'', \cdots \}$ with commuting scalar values. Let $$\dot{X} =
\nabla X = JDX = J(X'-X)/\tau$$ where $\tau$ is an elementary time
step (If $X$ denotes a times series value at time $t$, then $X'$
denotes the value of the series at time $t + \tau.$). The shift
operator $J$ is defined by the equation $XJ = JX'$ where this
refers to any point in the time series so that $X^{(n)}J =
JX^{(n+1)}$ for any non-negative integer $n.$ Moving $J$ across a
variable from left to right, corresponds to one tick of the clock.
This discrete, non-commutative time derivative satisfies the
Leibniz rule. \bigbreak

This derivative $\nabla$ also fits a significant pattern of
discrete observation. Consider the act of observing $X$ at a given
time and the act of observing (or obtaining) $DX$ at a given time.
Since $X$ and $X'$ are ingredients in computing $(X'-X)/\tau,$ the
numerical value associated with $DX,$ it is necessary  to let the
clock tick once, Thus, if one first observe $X$ and then obtains
$DX,$ the result is different (for the $X$ measurement) if one
first obtains $DX,$ and then observes $X.$ In the second case, one
finds the value $X'$ instead of the value $X,$ due to the tick of
the clock. \bigbreak

\begin{enumerate}
\item Let $\dot{X}X$ denote the sequence: observe $X$, then obtain
$\dot{X}.$ \item Let $X\dot{X}$ denote the sequence: obtain
$\dot{X}$, then observe $X.$
\end{enumerate}
\bigbreak

The commutator $[X, \dot{X}]$ expresses the difference between
these two orders of discrete measurement. In the simplest case,
where the elements of the time series are commuting scalars, one
has
$$[X,\dot{X}] = X\dot{X} - \dot{X}X =J(X'-X)^{2}/\tau.$$
Thus one can interpret the equation $$[X,\dot{X}] = Jk$$ ($k$ a
constant scalar) as $$(X'-X)^{2}/\tau = k.$$ This means that the
process is a walk with spatial step $$\Delta = \pm \sqrt{k\tau}$$
where $k$ is a constant. In other words, one has the equation
$$k = \Delta^{2}/\tau.$$
This is the diffusion constant for a Brownian walk. A walk with
spatial step size  $\Delta$ and time step $\tau$ will satisfy the
commutator equation above exactly when the square of the spatial
step divided by the time step remains constant. This shows that
the diffusion constant of a Brownian process is a structural
property of that process, independent of considerations of
probability and continuum limits. \bigbreak

\noindent {\bf Heisenberg/Schr\"{o}dinger Equation.} Here is how
the Heisenberg form of Schr\"{o}dinger's equation fits in this
context. Let the time shift operator be given by the equation
$J=(1 + H\Delta t/i \hbar).$ Then the non-commutative version of
the discrete time derivative is expressed by the commutator
$$\nabla\psi = [\psi, J/\Delta t],$$ and we calculate
$$\nabla \psi = \psi[(1 + H \Delta t/i \hbar)/\Delta t] -
[(1 +  H\Delta t/i \hbar)/\Delta t] \psi = [\psi, H]/i \hbar,$$
$$i \hbar \nabla \psi = [\psi, H].$$
This is exactly the Heisenberg version of the Schr\"{o}dinger
equation. \bigbreak

\noindent {\bf Dynamics and Gauge Theory.} One can take the
general dynamical equation in the form
$$dX_{i}/dt = {\cal G}_{i}$$ where $\{ {\cal G}_{1},\cdots, {\cal G}_{d} \}$
is a collection of elements of $\cal A.$ Write ${\cal G}_{i}$
relative to the flat coordinates via ${\cal G}_{i} = P_{i} -
A_{i}.$ This is a definition of $A_{i}$ and $\partial F/\partial
X_{i} = [F,P_{i}].$ The formalism of gauge theory appears
naturally. In particular, if $$\nabla_{i}(F) = [F, {\cal
G}_{i}],$$ then one has the curvature $$[\nabla_{i}, \nabla_{j}]F
= [R_{ij}, F]$$ and
$$R_{ij} = \partial_{i} A_{j} - \partial_{j} A_{i} + [A_{i}, A_{j}].$$  This is the well-known formula for the curvature of a gauge
connection. Aspects of geometry arise naturally in this context,
including the Levi-Civita connection (which is seen as a
consequence of the Jacobi identity in an appropriate
non-commutative world). \bigbreak

One can consider the consequences of the commutator $[X_{i},
\dot{X_{j}}] = g_{ij}$, deriving that
$$\ddot{X_{r}} = G_{r} + F_{rs}\dot{X^{s}} + \Gamma_{rst}\dot{X^{s}}\dot{X^{t}},$$
where $G_{r}$ is the analogue of a scalar field, $F_{rs}$ is the
analogue of a gauge field and $\Gamma_{rst}$ is the Levi-Civita
connection associated with $g_{ij}.$ This decompositon of the
acceleration is uniquely determined by the given framework.
\bigbreak

\section{Spacetime and information like a fiber space}

We present a toy model where in every point of time exist an
information, connected with the spacetime variation.

We shall call $E$ with elements $z^{A}$ in E, which are:
\begin{equation}
 (x^{\mu},\theta^{\alpha}) \ ; .                           \label{1}
\end{equation}

The fiber space is $E(V,W,SU(2))$, where the basis $V$ is
spacetime a fiber $W$ is a information qubit. $SU(2)$ is a Lie
group acting on the fiber and E is ¨the cartesian product of V and
W.

On E we define one forms as usually:
\begin{eqnarray}
 \omega^{\mu} & = & \Omega^{\mu}+
      i\bar{\omega}^{\alpha}(\gamma^{\mu})_{\alpha\beta}\,\theta^{\beta}
                            \ , \label{2} \\
 \omega^{\alpha} & = & d \theta^{\alpha} \ .     \label{3}
\end{eqnarray}

For one discrete time dimension we have $\theta^{\alpha}$ discrete
time series.

For \(\Omega^{\mu}=dx^{\mu}\) is valid that
$\Omega_{\mu}\Omega^{\mu}$ is  $SU(2)$ invariant.

From fiber structure is known that $\Omega^{\mu}$ is connected
with arbitrary form $\omega^{\mu}$ on $E$ via following way:
\begin{equation}
 \omega^{\mu} = \Omega^{\mu}+
      \bar{\omega}^{\alpha}\Gamma_{\alpha}^{\mu} \ , \label{2'}
\end{equation}
where $\Gamma_{\alpha}^{\mu}$ are connection forms on E and
$\Gamma$ are Pauli matrices.

This forms define transformation from one fiber to another when
infinitesimal changes in the base are realized.

On in $W$ is one-form $\Omega^{\mu}$ and on $V$  form
$\omega^{\alpha}$. From $x_\mu=\bar{\theta}'\gamma^{\mu} \theta $
follows $x^{\mu}$ is $\theta^{\alpha}$ and $\Omega^{\mu}$ has the
form:
\begin{eqnarray*}
 \Omega^{\mu} & = & dx^{\mu}           \\
              & = & \left(\frac{\partial x^{\mu}}{\partial\theta^{\alpha}}
              \right) d\theta^{\alpha}         \\
              & = & \left(\frac{\partial x^{\mu}}{\partial\theta^{\alpha}}
              \right) \omega^{\alpha}  \ .
\end{eqnarray*}

It means that $\Omega^{\mu}$ on $W$ is given via $\omega^{\alpha}$
on $V$.

It is valid: \( \frac{\partial x^{\mu}}{\partial \theta^{\alpha}}=
    \bar{\theta}'^{\beta}(\gamma^{\mu})_{\beta\alpha}\), \\
and we get: \(\Omega^{\mu}=\bar{\theta}'(\gamma^{\mu})\omega\).

Arbitrary object $Y^{J}$ in E transforms as:
\begin{equation}
 d Y^{J}+Y_{\mu}^{J}\omega^{\mu} = \bar{\omega}^{\alpha}Y_{\alpha}^{J} \ .
                      \label{4}
\end{equation}

Covariant derivation follows as
\begin{equation}
 \nabla \Phi(x,\theta) = d\Phi(x,\theta)+\Phi_{\mu}(x,\theta)+\Omega^{\mu}
        \ .              \label{4'}
\end{equation}

So we get:
\[ \Omega^{\mu} = \omega^{\mu}-\bar{\omega}^{\alpha}\Gamma_{\alpha}^{\mu} \]
and then:
\begin{equation}
 \nabla \Phi(x,\theta) = d\Phi(x,\theta)+\Phi_{\mu}(x,\theta)
 (\omega^{\mu}-\bar{\omega}^{\alpha}\Gamma_{\alpha}^{\mu})
        \ .              \label{5}
\end{equation}

We get:
\[ \Phi_{\mu}(x,\theta)\omega^{\mu}=
    \bar{\omega}^{\alpha}\Phi_{\alpha}(x,\theta)-d\Phi(x,\theta) \]
and following:
\begin{eqnarray}
  \nabla \Phi(x,\theta) & = & \bar{\omega}^{\alpha}\Phi_{\alpha}(x,\theta)
      -\Phi_{\mu}(x,\theta)\,\bar{\omega}^{\alpha}\Gamma_{\alpha}^{\mu}
      \nonumber     \\
        & = & \bar{\omega}^{\alpha}(\Phi_{\alpha}(x,\theta)
      -\Phi_{\mu}(x,\theta)\,\Gamma_{\alpha}^{\mu})      \label{6}     \\
        & = & \bar{\omega}^{\alpha}\Phi_{i^{\alpha}}(x,\theta) \ .
      \nonumber
\end{eqnarray}

It is valid:
\[ \Phi_{i^{\alpha}}(x,\theta)=\partial_{\alpha}\Phi(x,\theta)
   - i(\gamma^{\mu}\theta)_{\alpha}\,\partial_{\mu}\Phi(x,\theta)=
   D_{\alpha}\Phi(x,\theta) \ , \]
where \( D_{\alpha}=\partial_{\alpha}
    -i(\gamma^{\mu}\theta)_{\alpha}\,\partial_{\mu}=\partial_{\alpha}
    -\Gamma_{\alpha}^{\mu}\,\partial_{\mu} \)
is the covariant derivation.

\section{Conclusions}

 Here we show another aspect of the connection of the quantum time
 and quantum information. We show discrete nature of quantum time
 and present the idea of the nontrivial connection of quantum information and
 spacetime via fiber space.

   This work was supported by Grant T300100403 GA AV CR  .


\bigskip
\begin{thebibliography}{2}
\bibitem{1} J. Hruby, ``Supersymmetry and qubit field theory'',,
            arXiv: quant-ph/0402188,(2004),
            J.Hruby,``A role of topology and quantum information in physics'',,
            arXiv: quant-ph/0502118,(2005).

\bibitem{2} L.H. Kaufmann,``Non-Commutative Worlds- A Summary'',,
            arXiv: quant-ph/0503198,v.2(2005)
\end{thebibliography}
\end{document}